\documentclass[aps,prx,twocolumn,superscriptaddress,showpacs,floatfix,longbibliography]{revtex4-2}
\usepackage{amsmath,amssymb,wasysym,graphicx}
\pdfoutput=1
\usepackage[utf8]{inputenc}
\usepackage[T1]{fontenc}
\usepackage{xcolor}

\IfFileExists{newtxtext.sty}
   {\usepackage{newtxtext,newtxmath}}
   {\IfFileExists{stix.sty}
      {\usepackage{stix}}
      {\IfFileExists{mathptmx.sty}
      {\usepackage{mathptmx}}{} } }

\usepackage{textcomp}

\usepackage{bm}

\IfFileExists{siunitx.sty}{\usepackage{booktabs,siunitx}}{}

\usepackage{color}
\definecolor{LinkColor}{rgb}{0.256,0.439,0.588}
\usepackage{hyperref}
\hypersetup{
   colorlinks=true,
   citecolor=LinkColor,
   linkcolor=LinkColor,
   urlcolor=LinkColor
}

\usepackage{pifont}
\usepackage[normalem]{ulem}

\usepackage{multirow}

\begin{document}

\title{Dirac fermions with plaquette interactions. III. $SU(N)$ phase diagram \\
with Gross-Neveu criticality and first-order phase transition}

\author{Yuan Da Liao}
\affiliation{State Key Laboratory of Surface Physics, Fudan University, Shanghai 200438, China}
\affiliation{Center for Field Theory and Particle Physics, Department of Physics, Fudan University, Shanghai 200433, China}

\author{Xiao Yan Xu}
\email{xiaoyanxu@sjtu.edu.cn}
\affiliation{Key Laboratory of Artificial Structures and Quantum Control (Ministry of Education), School of Physics and Astronomy, Shanghai Jiao Tong University, Shanghai 200240, China}

\author{Zi Yang Meng}
\email{zymeng@hku.hk}
\affiliation{Department of Physics and HKU-UCAS Joint Institute of Theoretical and Computational Physics, The University of Hong Kong, Pokfulam Road, Hong Kong SAR, China}

\author{Yang Qi}
\email{qiyang@fudan.edu.cn}
\affiliation{State Key Laboratory of Surface Physics, Fudan University, Shanghai 200438, China}
\affiliation{Center for Field Theory and Particle Physics, Department of Physics, Fudan University, Shanghai 200433, China}
\affiliation{Collaborative Innovation Center of Advanced Microstructures, Nanjing 210093, China}

\begin{abstract}
Inspired by the recent works~\cite{liaoGross2022,liaoGrossSU42022} of $SU(2)$ and $SU(4)$ Dirac fermions subjected to plaquette interactions on square lattice, here we extend the large-scale quantum Monte Carlo investigations to the phase digram of correlated Dirac fermions with $SU(6)$ and $SU(8)$ symmetries subjected to the plaquette interaction on the same lattice. From $SU(2)$ to $SU(8)$, the rich phase diagram exhibits a plethora of emerging quantum phases such as the Dirac semimetal, the antiferromagnetic Mott insulator, valence bond solid (VBS) and the Dirac spin liquid and phase transitions including the Gross-Neveu chiral transitions with emergent continuous symmetry, the deconfined quantum criticality and the first order transition between interaction-driven columnar VBS and plaquette VBS. These rich phenomena coming from the simple-looking lattice models, firmly convey the message that the interplay between the $SU(N)$ Dirac fermions -- with enhanced internal symmetries -- and extended plaquette interactions -- beyond the on-site Hubbard type -- is the new playground to synthesise novel highly entangled quantum matter both at the model level and with experimental feasibilities.
\end{abstract}

\date{\today}

\maketitle
\section{Introduction}
Recently, the interests of interacting Dirac fermions have been rekindled by the findings that one can extend the interactions from the traditional on-site Hubbard repulsion~\cite{sorellaSemi1992,herbutInteractions2006,mengQuantum2010,changQuantum2012,otsukaUniversal2016,toldinFermionic2015,ouyangProjection2021} to the more extended ones that involve all sites in a plaquette and solve the problem with unbiased large-scale quantum Monte Carlo (QMC) simulations~\cite{liaoGross2022,liaoGrossSU42022,ouyangProjection2021}. Such extension is also coincided with the great reach trend in the 2D quantum (moir\'e) materials such as twisted bilayer graphene (TBG) and twisted transition metal dichalcogenides (TMD), which feature the high tunability by twisting angles, gating and dielectric environment and the perfect 2D setting with flat-bands~\cite{tramblyLocalization2010,tramblyNumerical2012,
bistritzerMoire2011,Santos2012,lopesGraphene2007,caoUnconventional2018,
shenCorrelated2020,xieSpectroscopic2019,KhalafCharged2021,KevinStrongly2020,
pierceUnconventional2021,caoCorrelated2018,xuKekule2018,liaoValence2019,
liaoCorrelated2021,luSuperconductors2019,moriyamaObservation2019,sharpeEmergent2019,serlinIntrinsic2020,chenTunable2020,
rozhkovElectronic2016,ChatterjeeSkyrmion2020,kerelskyMaximized2019,rozenEntropic2021,liQuantum2021,
tomarkenElectronic2019,soejimaEfficient2020,liuSpectroscopy2021,KhalafSoftmodes2020,
ZondinerCascade2020,saitoPomeranchuk2021,GhiottoCriticality2021,SchindlerTrion2022,WangTMD2020,
Parkchern2021,liaoCorrelation2021,anInteraction2020,huangGiant2020,liLattice2021,panSport2022,
panThermodynamic2022,zhangSuperconductivity2021,panDynamical2022,zhangMomentum2021}; as well as in the kagome transition metal intermetallic comounds with both flat-bands and Dirac cones depending on the chemical composition, such as Ni$_3$In, FeSn and many others~\cite{xieSpin2021,kangDirac2020,yeflat2021}, that the interplay of the quantum geometry of Dirac fermion wavefunctions and the strong extended or long-range Coulomb interactions, gives rise to a plethora of correlated phases, emerged and has been continuously emerging in an astonishing pace. 

These theoretical and experimental progresses all point towards the importance and the crying need to understand the effect of extended interaction on the band structures with non-trivial quantum metric in their wavefunctions, realized either in the form of non-trivial topological numbers (such as Chern number~\cite{sharpeEmergent2019,serlinIntrinsic2020,liQuantum2021}) or incarnated in the various realizations of Dirac cones.

At the model level, the investigations of the Dirac fermions, either come from the purely theoretical construction or the graphene and moir\'e lattice models, have been successful and fruitful~\cite{sorellaSemi1992,herbutInteractions2006,mengQuantum2010,changQuantum2012,otsukaUniversal2016,toldinFermionic2015,ouyangProjection2021,liaoValence2019,xuTopological2017,xuKekule2018,liaoCorrelated2021,liaoCorrelation2021,liaoGross2022,liaoGrossSU42022,heDynamical2018,liuDesigner2020,zhuQuantum2022}. In our previous works, with $SU(2)$ and $SU(4)$ Dirac fermions subjected to plaquette interactions on square lattice, the interaction-driven Gross-Neveu (GN) chiral XY transition, the deconfined quantum criticality and the Dirac spin liquid phase with emergent Dirac spinon coupled with gauge fields, have been discovered via large-scale quantum Monte Carlo (QMC) simulations and the systematic finite-size scaling analyses~\cite{liaoGross2022,liaoGrossSU42022}. Here, we extend the investigation in a more systematic manner and explore the entire phase diagram of the $SU(N)$ with $N=2,4,6,8$ correlated Dirac fermions on the square lattice with plaquette interaction.

\begin{figure}[htp!]
\includegraphics[width=\columnwidth]{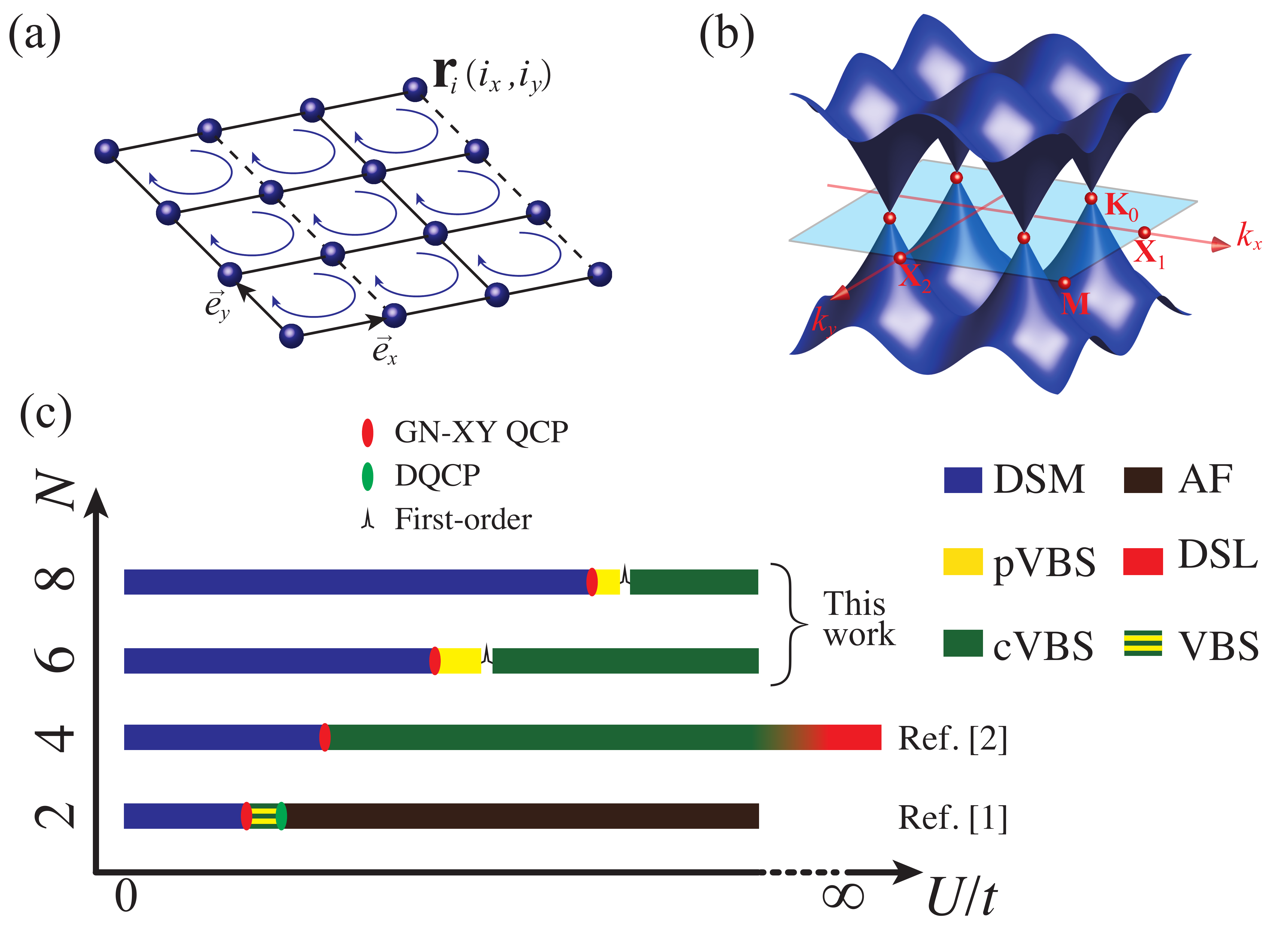}
\caption{(a) The $\pi$-flux square lattice realising Dirac fermions. The solid and dash lines correspond to hopping amplitude with $-t$ and $t$. The position of site $i$ is given as $\mathbf{r}_i = i_{x}\vec{e}_{x}+i_{y}\vec{e}_{y}$, with $\vec{e}_x$ and $\vec{e}_y$ the unit vectors along $x$ and $y$ directions. (b) The dispersion of free $\pi$-flux model in the first Brilliouin zone (BZ). The Dirac cones are located at $\mathbf{K}_0=(\pm \frac{\pi}{2},\pm \frac{\pi}{2})$. Other high symmetry points $\mathbf{\Gamma}=(0,0)$, $\mathbf{X}_1=(\pi,0)$, $\mathbf{X}_2=(0,\pi)$ and $\mathbf{M}=(\pi,\pi)$ are also denoted. (c) The phase diagram of $SU(N)$ Dirac fermions with plaquette interaction. DSM denotes  the Dirac semimetal, pVBS denotes the plaquette valence bond solid phase, cVBS denotes the columnar valence bond solid phase, AF denotes the antiferromagnetic Mott insulator, DSL denotes the Dirac spin liquid and VBS denote a valence solid state which is difficult to distinguish from pVBS and cVBS at $SU(2)$. The red ellipsoids represent the location of continuous QCP with Gross-Neveu chiral XY universality. The green ellipsoid represents the location of DQCP. The black pulse represents the position of first-order phase transition.}
	\label{fig:model}
\end{figure}

The rich phase diagram, shown in Fig.~\ref{fig:model}, is spanned by the axes of $N$ -- the fermion flavors and the $U$ -- the plaquette interaction strength. Besides the results of $N=2$ and $4$ cases~\cite{liaoGross2022,liaoGrossSU42022}, here we find at $N=6$ and $8$, there are new phases and transitions happened, namely, at small plaquette interaction, the $SU(6)$ and $SU(8)$ Dirac fermions stay intact, but as a function of $U$, they all experience GN quantum critical point (QCP) such that the Dirac fermions acquire mass and form bound state of plaquette valence bond solid (pVBS) with lattice translational symmetry breaking. The GN transition belongs to the GN-XY type with an emergent U(1) symmetry in the valence bond solid (VBS) order parameter histogram, similar with the situation in their $SU(2)$ and $SU(4)$ cousins. In the pVBS phase, the electrons are not fully localized and can still resonate within one plaqutte.

However, further increasing the plaquette interaction $U$, the electrons which are resonating inside a plaquette in the pVBS further localized into coherent state among the nearest-neighbor bond and form the columnar VBS (cVBS) state.
We find the transition between the two VBS states is first order, consistent with the fact that they break different lattice symmetries:
the pVBS breaks lattice translation symmetry with a two-by-two unit cell, and the cVBS breaks translation with a one-by-two unit cell but further breaks the rotation symmetry.
Moreover, these two VBS phases can be elegantly distinguished from the rotation in the signals in their order parameter histogram from the QMC data. 
The further questions such as the $N\to\infty$ limit phase diagram and the possible experimental realisation of the our $SU(N)$ correlated Dirac fermion with plaquette interaction lattice model, are discussed as well.

The rest of the paper is organized as follows: in Sec.~\ref{sec:ii}, we explain the model and QMC method employed in this work. Then in Sec.~\ref{sec:iii}, the numerical results are given. We first briefly warmup the readers with our previous results in the $SU(2)$ and $SU(4)$ cases~\cite{liaoGross2022,liaoGrossSU42022}, and then focus on the $SU(6)$ and $SU(8)$ cases, with the GN-XY QCP and the pVBS-cVBS first order transition discussed in details, including the critical scaling analysis and the order parameter histograms. Sec.~\ref{sec:iv} summarizes the entire $SU(N)$ phase diagram obtained, and we reiterate the physical meanings of our model level discoveries towards the on-going experimental efforts in 2D quantum moir\'e materials such as TBG, TMD and kagome metals with Dirac fermions subjected to extended and long-range interactions.

\section{Model and Method}
\label{sec:ii}
\subsection{$SU(N)$ plaquette Hubbard model on $\pi$-flux square lattice}
We investigate the $SU(N)$ plaquette Hubbard model on $\pi$-flux square lattice
\begin{equation}
	H = \sum_{\langle i j\rangle, \alpha} t_{i j}\left(c_{i \alpha}^{\dagger} c_{j \alpha}+\text { H.c. }\right) + U \sum_{\square}\left(n_{\square}-N/2 \right)^{2},
\label{eq:eq1}
\end{equation}
where $c_{i \alpha}^{\dagger}$ and $c_{i \alpha}$ represent creation and annihilation operators for fermions on site $i$ with flavor indices $\alpha\in [1,N]$ with $SU(N)$ symmetry,
$\langle i j\rangle$ represent the first nearest neighbors.
Since every site is shared by $4$ plaquettes, we define the extended particle number operator at each $\square$-plaquette as $n_{\square}  \equiv \frac{1}{4} \sum_{i \in \square} n_{i} $ with $n_{i}=\sum_{\alpha=1}^{N} c_{i \alpha}^{\dagger} c_{i \alpha}$
and $\langle n_{\square} \rangle = N/2$ at half-filling.
$U$ is the tunable repulsive plaquette interaction strength.

As shown in Fig.~\ref{fig:model}(a), a solid bond denotes hopping amplitude $t$, a dash one denotes $-t$, i.e., $t_{i,i+\vec{e}_{x}} = t$ and $t_{i,i+\vec{e}_{y}} = (-1)^{i_x} t$. The position of site $i$ is given as $\mathbf{r}_i = i_{x}\vec{e}_{x}+i_{y}\vec{e}_{y}$. 
And we set the energy unit $t=1$ throughout this paper.
Such convention bestows a $\pi$-flux in each $\square$-plaquette and gives rise to the dispersion relation $\epsilon(\mathbf{k}) = \pm 2 t \sqrt{\cos^2 k_x + \cos^2 k_y} $ shown in Fig.~\ref{fig:model}(b).
It's easy to find that there are gapless Dirac cones located at momentum point $\mathbf{K}_0=\left(\pm \frac{\pi}{2}, \pm\frac{\pi}{2}\right)$, which indicates our model is in the Dirac semimetal (DSM) phase at zero temperature when $U/t=0$.
The distances between these gapless Dirac cones won't change in the Brillouin zone (BZ), no matter how we chose the gauge~\cite{liaoGross2022}, thus we can perform our numerical calculations in the first BZ of the original square lattice.
We also denote other high symmetry points $\mathbf{\Gamma}$, $\mathbf{X}_1$, $\mathbf{X}_2$ and $\mathbf{M}$ for discussing further results conveniently. As for the extended interaction term in Eq.~\eqref{eq:eq1}, since it contains the onsite, first and second nearest neighbor repulsions in one plaquette, we dub it the plaquette interaction. 

\subsection{Projector Quantum Monte Carlo Method}
We use the projector quantum Monte Carlo (PQMC) method to investigate the ground states and the quantum phase transitions between them.
With the help of particle-hole symmetry at half-filling, our PQMC simulation won't suffer from sign problem~\cite{wuSufficient2005,xuKekule2018,liaoValence2019,liaoCorrelation2021,Xu2022quantum,panSign2022}.
In PQMC method, one obtains the ground state wave function $\vert \Psi_0 \rangle$ via the projector of a trial wave function $\vert \Psi_T \rangle$ as, $\vert \Psi_0 \rangle \equiv \lim\limits_{\Theta \to \infty} e^{-\frac{\Theta}{2} H} \vert \Psi_T \rangle$. 
A physical observable $\hat{O}$ can be correspondingly evaluated as 
\begin{equation}
\label{eq:observablepqmc}
\langle \hat{O} \rangle = \frac{\langle \Psi_0 \vert \hat{O} \vert \Psi_0 \rangle}{\langle \Psi_0 \vert \Psi_0 \rangle} 
						= \lim\limits_{\Theta \to \infty} \frac{\langle \Psi_T \vert  e^{-\frac{\Theta}{2} H} \hat{O}  e^{-\frac{\Theta}{2} H} \vert \Psi_T \rangle}{\langle \Psi_T \vert  e^{-\Theta H} \vert \Psi_T \rangle} ,
\end{equation}
where $H$ is the Hamiltonian and $\Theta$ the projection length. As in Eq.~\eqref{eq:eq1}, $H$ is consisted of the non-interacting $K\equiv \sum_{\langle i j\rangle, \alpha} t_{i j}\left(c_{i \alpha}^{\dagger} c_{j \alpha}+\text { H.c. }\right)$ and interacting $V\equiv U \sum_{\square}\left(n_{\square}-N/2 \right)^{2}$ parts that are usually not commute, we perform the Trotter decomposition to discretize $\Theta$ into $L_\tau$ slices, each slice has the thickness $\Delta\tau$ ($\Theta=L_\tau \Delta\tau$). Then
\begin{equation}
\langle\Psi_{T}|e^{-\Theta H}|\Psi_{T}\rangle=\langle\Psi_{T}|\left(e^{-\Delta\tau V }e^{-\Delta\tau K}\right)^{L_\tau}|\Psi_{T}\rangle+\mathcal{O}(\Delta{\tau}^{2}),
\end{equation}
after which, the non-interacting and interacting parts of the Hamiltonian are separated. 
Since the Trotter decomposition gives rise to a small systematic error $\mathcal{O}(\Delta\tau^2)$, we should set $\Delta\tau$ as a small number to achieve reliable results within the statistical errorbars of the Monte Carlo sampling. The interaction part contains quartic fermionic operators that can not be evaluated directly. 
One need to employ a $SU(N)$ symmetric Hubbard-Stratonovich (HS) decomposition, then the auxiliary fields will couple to the charge density.
For example, for our extended interaction, the HS decomposition is
\begin{equation}
e^{-\Delta\tau U(n_{\square}-N)^{2}}=\frac{1}{4}\sum_{\{s_{\square}\}}\gamma(s_{\square})e^{\alpha\eta(s_{\square})\left(n_{\square}-N\right)}
\label{eq:decompo}
\end{equation}
with $\alpha=\sqrt{-\Delta\tau U}$, $\gamma(\pm1)=1+\sqrt{6}/3$,
$\gamma(\pm2)=1-\sqrt{6}/3$, $\eta(\pm1)=\pm\sqrt{2(3-\sqrt{6})}$,
$\eta(\pm2)=\pm\sqrt{2(3+\sqrt{6})}$ and the sum is taken over the auxiliary fields $s_{\square}$ on each $\square$-plaquette. We then arrive at the following formula with constant factors omitted
\begin{footnotesize}
\begin{equation}
\langle\Psi_{T}|e^{-\Theta H}|\Psi_{T}\rangle = \sum_{\{s_{\square,\tau}\}} \prod_{\square,\tau}\gamma(s_{\square,\tau})e^{-N\alpha\eta(s_{\square,\tau})} \det\left[P^{\dagger}B(\Theta,0)P\right],
\label{eq:mcweight}
\end{equation}
\end{footnotesize}
where $P$ is the coefficient matrix of trial wave function $|\Psi_T\rangle$; $B$ is a matrix defined as $B(\tau+1,\tau)=e^{V_\tau}e^{-\Delta\tau K}$ 
, here $V_\tau$ is the corresponding matrix representation at time slice $\tau$, and $B$ has a mathematical property $B(\tau_3,\tau_1)=B(\tau_3,\tau_2)B(\tau_2,\tau_1)$.
In practice, we use the ground state wavefunction of the half-filled non-interacting part $K$ 
as the trial wave function. The measurements are performed near $\tau=\Theta/2$.
The Metropolis updates of the auxiliary fields are further performed based on the weight defined in the sum of Eq.~\eqref{eq:mcweight}.  
Single-particle (fermion bilinear) observables are measured via Green's function directly and the correlation functions of collective excitations are measured from the products of single-particle Green's function based on their corresponding form after Wick-decomposition for each auxiliary field configuration and averaged over to have their ensemble means and variances. 
The equal time Green's function are calculated as
\begin{equation}
G(\tau,\tau)=1-R(\tau)\left(L(\tau)R(\tau)\right)^{-1}L(\tau)
\end{equation}
with $R(\tau)=B(\tau,0)P$, $L(\tau)=P^{\dagger}B(\Theta,\tau)$. 
The imaginary-time displaced Green's function  
$G(\tau,0) \equiv \left\langle \mathbf{c} \left(\frac{\tau}{2}\right) \mathbf{c}^{\dagger}\left(-\frac{\tau}{2}\right)\right\rangle$
are calculated as
\begin{equation}\label{eq:dynamic-greenfunction}
\left\langle \mathbf{c} \left(\frac{\tau}{2}\right) \mathbf{c}^{\dagger}\left(-\frac{\tau}{2}\right)\right\rangle=\frac{\left\langle\Psi_{T}\left|e^{-\left(\frac{\Theta}{2}+\frac{\tau}{2}\right) H} \mathbf{c} e^{-\tau H} \mathbf{c}^{\dagger} e^{-\left(\frac{\Theta}{2}-\frac{\tau}{2}\right) H}\right| \Psi_{T}\right\rangle}{\left\langle\Psi_{T}\left|e^{-\Theta H}\right| \Psi_{T}\right\rangle},
\end{equation}
where $\mathbf{c} = \{c_1, c_2 \cdots c_{N_s} \}$, $N_s=L\times L$ is the system size.
More technique details of PQMC method can be found in  Refs.~\cite{assaadWorld-line2008,xuKekule2018,liaoValence2019,liaoCorrelation2021}. 

\section{Results}
\label{sec:iii}
\subsection{Phase diagram}
The phase diagram of our model spanned by the axes of $N$ and the interaction strength $U$ is schematically shown in Fig.~\ref{fig:model} (c), and it can be seen that there are richer and more  complicated phases and transitions than the similar $SU(N)$ phase diagram of Dirac fermion subject to the on-site Hubbard interaction~\cite{toldinFermionic2015,otsukaUniversal2016,zhouMott2018,ouyangProjection2021}.
For $N=2$, in our phase diagram, as increasing $U$ the system will transit from Dirac semimetal (DSM) to a VBS state through a Gross-Neveu chiral XY continuous phase transition (the VBS at $SU(2)$ case is robust but weak, and difficult to distinguish from pVBS and cVBS), then transit from VBS to antiferromagnetic (AF) Mott insulator state through a deconfined quantum critical phase transition~\cite{liaoGross2022}.
For $N=4$, as increasing $U$, the system will transit from DSM to a cVBS state through a Gross-Neveu chiral XY continuous phase transition, then transit to a possible U(1) Dirac spin liquid (QSL) at the $U=\infty$ yet tractable interaction limit~\cite{liaoGrossSU42022,ouyangProjection2021}, with continuum spinon spectrum and power-law magnetic susceptibility in temperature, consistent with theoretical predictions for Dirac spin liquid~\cite{ranProjected2007}.
For $N=6$ and $8$, the corresponding phase diagrams have a similar structure. As increasing $U$, the system will transit from  DSM to a pVBS state through a Gross-Neveu chiral XY continuous phase transition, then transit from pVBS to cVBS through a first-order phase transition.

The $SU(2)$ and $SU(4)$ the ground states and the phase transitions have been investigated in our previous studies~\cite{liaoGross2022,liaoGrossSU42022}. In this paper, we present the phase diagrams of $N=6$ and $8$ cases and summarize the entire $SU(N)$ phase diagram with the complete knowledge from all these efforts.

\subsection{Physical observables}
To detect the phase transitions, we computed several representative physical observables, and introduce them here before showing the data.
%
To locate the VBS order, we can define the VBS structure factor as
\begin{equation}
C_\text{VBS}^{\mathbf{e}} (\mathbf{k}, L) = \frac{1}{L^4} \sum_{i,j} e^{i \mathbf{k} \cdot\left(\mathbf{r}_{i}-\mathbf{r}_{j}\right)} \left\langle B^{\mathbf{e}}_{i} B^{\mathbf{e}}_{j},\right\rangle
\end{equation}
where $ B^{\mathbf{e}}_{i} = \frac{1}{N} \sum_{\alpha=1}^{N}\left( t_{i, i+\mathbf{e}}  c_{i, \alpha}^{\dagger} c_{i+\mathbf{e}, \alpha}+ \text {H.c.} \right)$ are gauge invariant bond operators with $\mathbf{e}=\vec{e}_x$ or $\vec{e}_y$.
For VBS order, $C_\text{VBS}^{\vec{e}_x}(\mathbf{k}, L)$ is peaked at momentum $\mathbf{X}_1$ and $C_\text{VBS}^{\vec{e}_y}(\mathbf{k}, L)$ is peaked at momentum $\mathbf{X}_2$.
It is known that the perfect VBS order on square lattice has $Z_4$ degeneracy, so $C_\text{VBS}^{\vec{e}_x}(\mathbf{X}_1, L)$ and $C_\text{VBS}^{\vec{e}_y}(\mathbf{X}_2, L)$ should be equivalent in ideal QMC simulations.
In view of that, we could define the square of the VBS order parameter as
\begin{equation}
 C_\text{VBS}(L) \equiv C_\text{VBS}^{\vec{e}_x}(\mathbf{X}_1, L)+C_\text{VBS}^{\vec{e}_y}(\mathbf{X}_2, L).
 \label{eq:eq10}
\end{equation}
From the VBS structure factor, one can further define the VBS correlation ratio $R_\text{VBS}$ as
\begin{equation}
R_\text{VBS}(U, L)=1-\frac{C_\text{VBS}^{\vec{e}_x}(\mathbf{X}_1+d \mathbf{q}, L)}{2C_\text{VBS}^{\vec{e}_x}(\mathbf{X}_1, L)}-\frac{C_\text{VBS}^{\vec{e}_y}(\mathbf{X}_2+d \mathbf{q}, L)}{2C_\text{VBS}^{\vec{e}_y}(\mathbf{X}_2, L)},
\end{equation}
where $|d\mathbf{q}| \sim 1/L$ is the smallest momentum on finite size lattice.
In principle, $R_\text{VBS}(U, L)$ goes to $1$ ($0$) for sufficiently large $L$ in the ordered (disordered) phases.
Near the QCP, $R_\text{VBS}$ will cross at one point for different $L$ and give rise to an estimate of the position of QCP ~\cite{campostriniFinite-size2014,kaulSpin2015,pujariInteraction2016,langQuantum2019}.

\begin{figure}[htpb!]
	\includegraphics[width=\columnwidth]{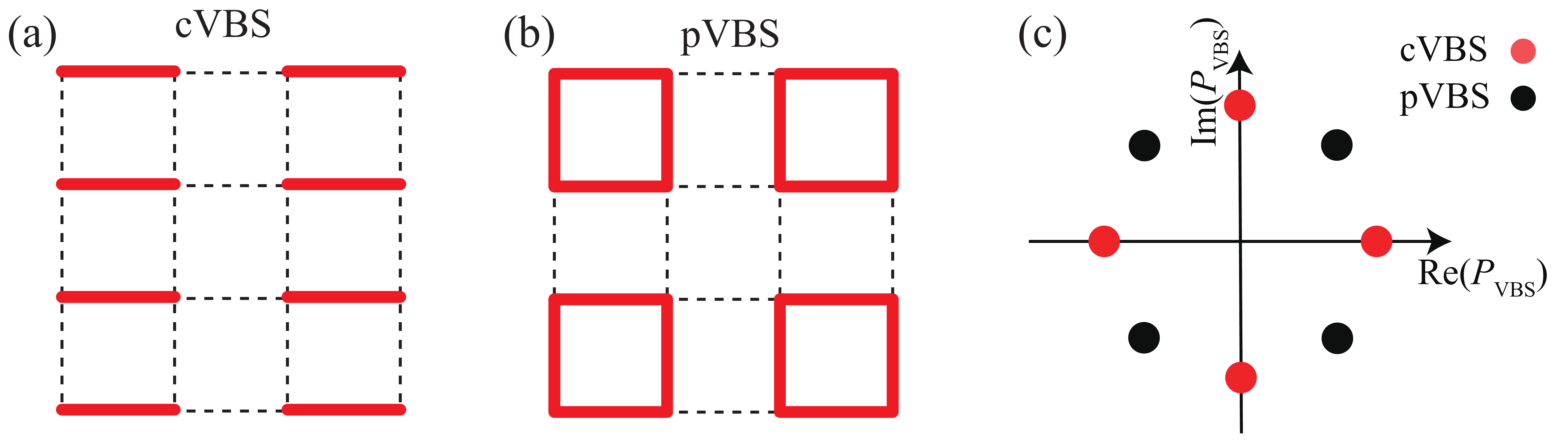}
	\caption{(a) and (b) are the symmetry breaking patterns of cVBS and pVBS. (c) The expected angular distribution of $\text{arg}(P_\text{VBS})$ for pVBS and cVBS.}
	\label{fig:VBS}
\end{figure}

Theoretically, as shown in Fig.~\ref{fig:VBS} (a) and (b), there are two kinds of VBS orders on square lattice, cVBS and pVBS, that could all give mass to the Dirac fermions and share the same order parameter $C_\text{VBS}$ in Eq.~\eqref{eq:eq10}.
%
To distinguish them, we could define an order parameter histogram as
\begin{equation}
\begin{aligned}
&P_{1}=\left(1 / L^{2}\right) \sum_{i}\left(B_{i}^{\vec{e}_{y}}+\omega B_{i}^{-\vec{e}_{x}}+\omega^{2} B_{i}^{-\vec{e}_{y}}+\omega^{3} B_{i}^{\vec{e}_{x}}\right) e^{i \mathbf{X}_{1} \cdot \mathbf{r}_{i}}, \\
&P_{2}=\left(1 / L^{2}\right) \sum_{i}\left(B_{i}^{\vec{e}_{y}}+\omega B_{i}^{-\vec{e}_{x}}+\omega^{2} B_{i}^{-\vec{e}_{y}}+\omega^{3} B_{i}^{\vec{e}_{x}}\right) e^{i \mathbf{X}_{2} \cdot \mathbf{r}_{i}}, \\
&P_{\text{VBS}}=P_{1}+P_{2}
\end{aligned}
\end{equation}
with $\omega=i$. 
For cVBS, the arguments of  $P_{\text{VBS}}$ are distributed at the angles $(0, \frac{\pi}{2}, \pi, \frac{3 \pi}{2})$,
while for pVBS at $(\frac{\pi}{4}, \frac{3 \pi}{4}, \frac{5 \pi}{4}, \frac{7 \pi}{4})$, as shown in Fig.~\ref{fig:VBS} (c).
Such definition of the order parameter histogram for VBS orders in interacting fermion systems is proposed and used in our previous study~\cite{liaoGrossSU42022,liaoValence2019}, which offers very sensitive probe to distinguish different two kinds of VBS orders.

We calculate the dynamical observables with the help of imaginary time-dependent Green's function $G_{ij}(\tau)\equiv  \langle c_i(\tau) c_j^\dagger (0) \rangle$  in Eq.~\eqref{eq:dynamic-greenfunction}.
In particular, the single-particle $\Delta_\text{sp}(\mathbf{k})$ at momentum $\mathbf{k}$ can be extracted from the fit of the imaginary time decay of 
\begin{equation}
G(\mathbf{k},\tau)=\left(1 / L^{4}\right) \sum_{i, j, \sigma} e^{i \mathbf{k} \cdot\left(\mathbf{r}_{i}-\mathbf{r}_{j}\right)} G_{ij}(\tau) \sim e^{-\Delta_{\text{sp}}(\mathbf{k}) \tau}.
\end{equation}
In the DSM phase, the single-particle gap at the Dirac cone, $\Delta_\text{sp}(\mathbf{K}_0)$ is zero. While in the Mott insulator phases such as VBS and AF, $\Delta_\text{sp}(\mathbf{K}_0)$ will give rise to a finite value in the thermodynamic limit (TDL).
We also compute the dynamical spin-spin correlation 
\begin{equation}
C(\mathbf{k}, \tau, L)=\frac{1}{L^{4}} \sum_{i, j} e^{i \mathbf{k} \cdot\left(\mathbf{r}_{i}-\mathbf{r}_{j}\right)} \sum_{\alpha\beta} \left\langle S_{\alpha\beta}(i,\tau) S_{\beta\alpha}(j,0)\right\rangle ,
\end{equation}
where $S_{\alpha \beta}(i,\tau)=c_{i, \alpha}^{\dagger}(0) c_{i, \beta}(\tau)-\frac{\delta_{\alpha \beta}}{2 N} \sum_{\gamma=1}^{2 N} c_{i, \gamma}^{\dagger}(0) c_{i, \gamma}(\tau)$,
the spin excitation gap $\Delta_\text{spin}(\mathbf{k})$ can be extracted from the relation $C(\mathbf{k}, \tau, L) \sim e^{-\Delta_{\text{spin}}(\mathbf{k}) \tau}$.
In the VBS phases, the spin gap is finite and in the AF phase, the spin excitation gap is zero due to the existence of Goldstone modes.

\subsection{Phase diagram for $SU(6)$: DSM-pVBS-cVBS}
We start the description of the data from the $SU(6)$ case. When $U$ is small, the system stays in the DSM phase due to the robustness of the Dirac cones. 
When interaction is strong enough, the Dirac cones will be gapped out and the system will transit into a Mott insulator, at where the corresponding QCP usually belongs to the Gross-Neveu universality~\cite{herbutTheory2009,herbutRelativistic2009,royMulticritical2011,royQuantum2013,royStrain-induced2014,royItinerant2018}.
Here, in our model, the Mott insulators are consisted of pVBS and cVBS orders as shown in Fig.~\ref{fig:VBS} (a) and (b).
We can monitor the VBS order with the help of $C_\text{VBS}$ and $R_\text{VBS}$ as a function of $U$ to determine the position of the QCP between DSM and VBS orders. 

\begin{figure}[htpb!]
\includegraphics[width=\columnwidth]{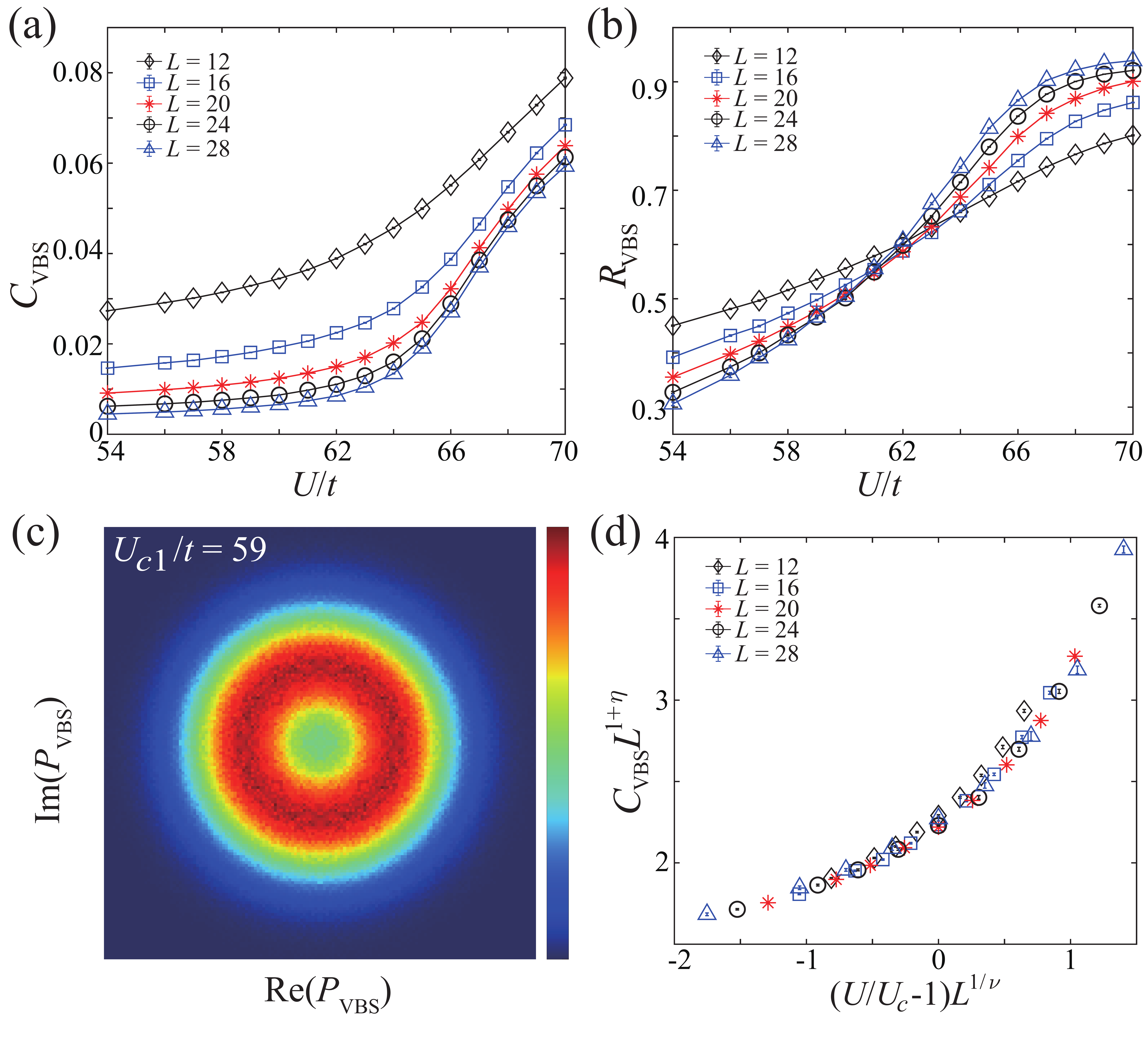}
\caption{The (a) $C_\text{VBS}$ and (b) $R_\text{VBS}$ for different $L$ as a function of $U$ for the $SU(6)$ case. The QCP is at  $U_{c1}/t=59(1)$. (c) The histogram of  $P_\text{VBS}$ at the GN-XY QCP ($U/t=59$, $L=24$)  shows a perfect emergent $U(1)$ symmetry. The color bar show the count of $P_\text{VBS}$ data points. (d) The data-collapse of $C_\text{VBS}$ in $SU(6)$ case with the GN chiral XY exponents $\nu=1.1(1)$ and $\eta=0.9(1)$ obtained.}
	\label{fig:cross-collapse}
\end{figure}

For the $SU(6)$ case, as shown in Fig.~\ref{fig:cross-collapse} (a), we find the $C_\text{VBS}$  gradually grows as a function of $U$, indicating that the VBS order (confirmed by the histogram discussed later as the pVBS) is developing continuously and there is a Gross-Neveu QCP separating the DSM and the pVBS order.
We can read the position of QCP from the crossing of $R_\text{VBS}$ at $U_{c1}/t=59(1)$, as shown in Fig.~\ref{fig:cross-collapse} (b).
At such QCP, the $Z_4$ discrete anisotropy of the pVBS order becomes irrelevant, and there emerges an $U(1)$ symmetry of the VBS order parameter~\cite{liaoGross2022,liaoGrossSU42022}. We plot the histogram of $\arg(P_\text{VBS})$ at $U/t=59$, as shown in Fig.~\ref{fig:cross-collapse} (c), and find there is indeed an emergent $U(1)$ symmetry, which suggests that the QCP between DSM and VBS order belongs to the (2+1)D GN chiral XY universality.  
What's more, we can further perform the data-collapse of $C_\text{VBS}$ according to the finite-size scaling relation $C_\text{VBS}=L^{-z-\eta} \mathcal{C}\left(L^{1 / v}\left(U/U_{c}-1 \right) \right)$ (assuming $z=1$), as shown in Fig.~\ref{fig:cross-collapse} (d), and obtain the critical exponents $\nu=1.1(1)$ and $\eta=0.9(1)$.
These critical exponents are consistent with previous numerical and theoretical study that investigate the same universality~\cite{rosensteinCritical1993,liFermion-induced2017}.

\begin{figure}[htpb!]
\includegraphics[width=\columnwidth]{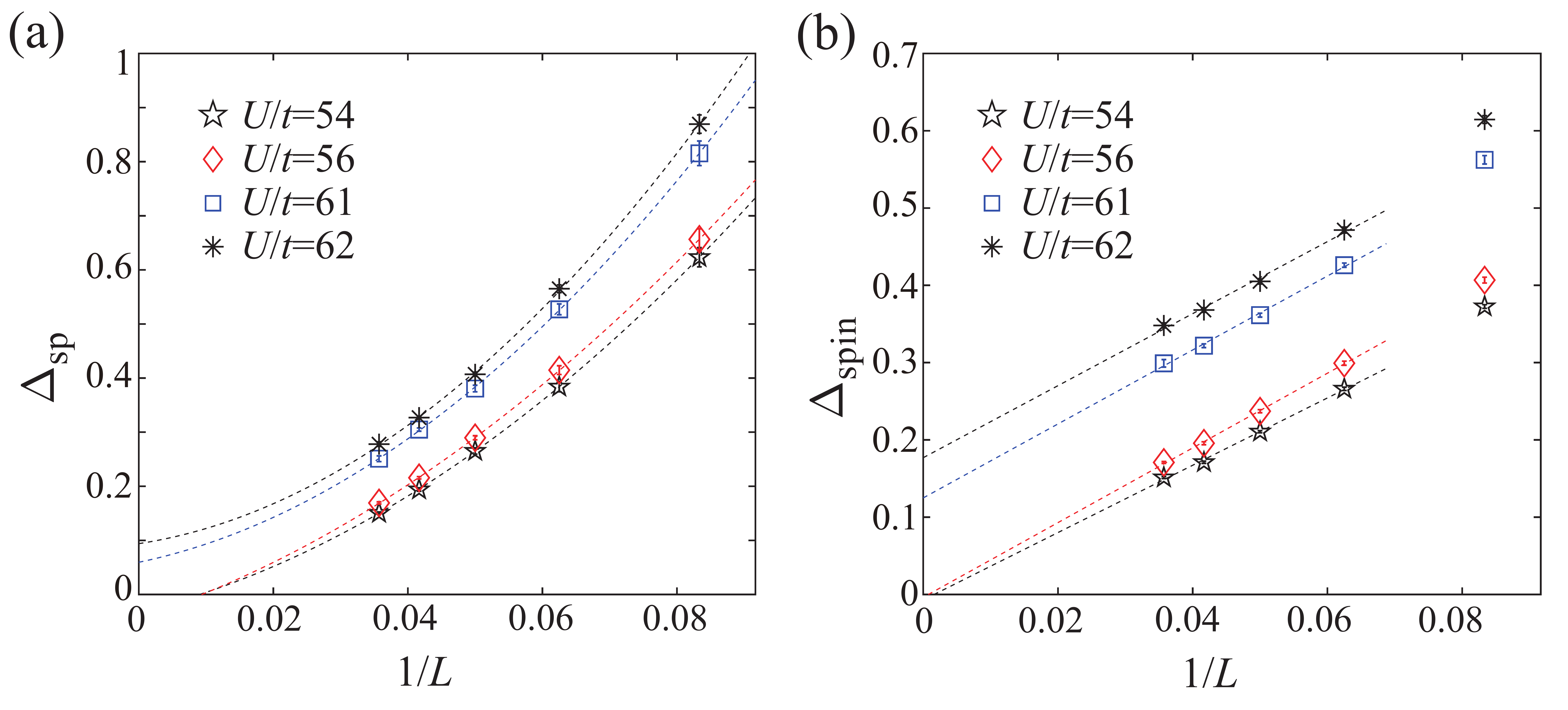}
\caption{The $1/L$ extrapolation of (a) $\Delta_\text{sp}$ and (b) $\Delta_\text{spin}$ in the $SU(6)$ case. 
We use 5 points, $L=12,16,20,24$ and $28$, to extrapolate $\Delta_\text{sp}$ and $\Delta_\text{spin}$ with a quadratic function.
Theoretically, $\Delta_\text{sp}$ and $\Delta_\text{spin}$ could be extrapolated to $0$ in DSM phase, while a finite value in an insulator phase. 
Thus the DSM-pVBS semimetal to Mott insulator transition happens between  $U/t=56$ and $U/t=61$, consistent with the results in Fig.~\ref{fig:cross-collapse}.}
\label{fig:dynamicSU6}
\end{figure}

The single-particle gap $\Delta_\text{sp}$ and spin excitation gap $\Delta_\text{spin}$ also open with the phase transition between the DSM phase and the insulating pVBS phase. As shown in Fig.~\ref{fig:dynamicSU6}, $\Delta_\text{sp}$ and $\Delta_\text{spin}$ go to $0$ as increasing $L$ to TDL for $U/t=54$ and $56$, which is consistent with the nature of DSM.
While, for $U/t=61$ and $62$, $\Delta_\text{sp}$ and $\Delta_\text{spin}$ can extrapolate to finite values, which means that the system stays in an insulator phase. These numerical results are consistent with the theoretical understanding of the GN chiral XY QCP separating the DSM and pVBS.
Moreover, we want to emphasize that the purpose of Fig.~\ref{fig:dynamicSU6} is not to determine the precise position of the GN-XY QCP, as it is difficult to perform finite size analysis with scaling function and the data collapse of the excitation gaps compared with that of the order parameter, which has been done in Fig.~\ref{fig:cross-collapse}. The purose of Fig.~\ref{fig:dynamicSU6} (a) and (b) is just to demonstrate that, after determining the precise position of $U_{c1}$, we can consistently see the DSM phase is gapless whereas the pVBS phase is gapped both in the single-particle and the spin channels.

\begin{figure}[htpb!]
\includegraphics[width=\columnwidth]{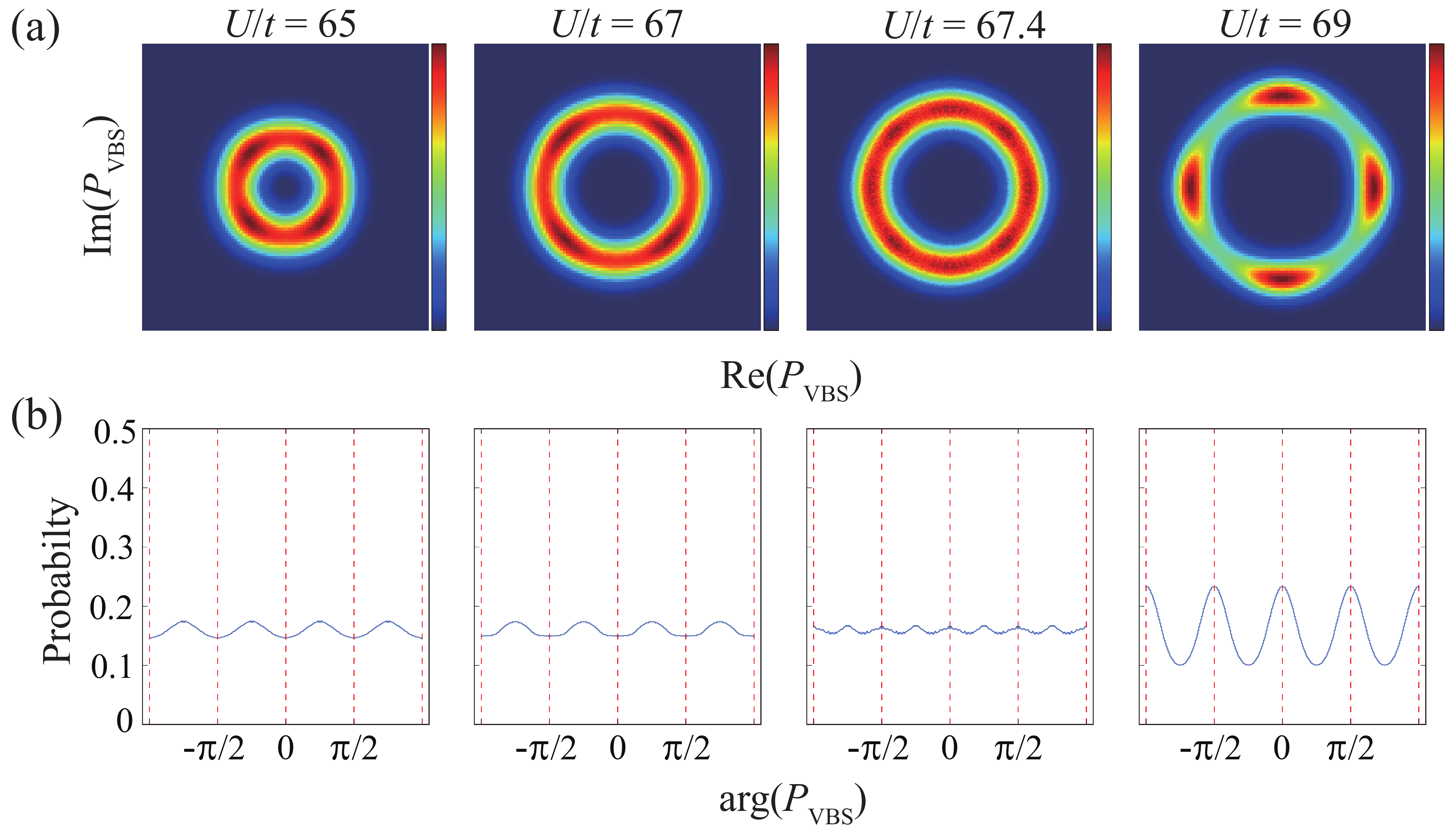}
\caption{(a) Histogram of $P_\text{VBS}$ and (b) the integrated distribution probability of arguments of $P_\text{VBS}$ in $SU(6)$ case. The color bars show the counts of $P_\text{VBS}$ data points. The transition from pVBS ($U/t=65,67$) and cVBS ($U/t=69$) is first order with phase coexistence ($U/t=67.4$) at the transition.}
	\label{fig:histogramSU6}
\end{figure}

More interestingly, further increasing $U$ to $U_{c2}/t \sim 67.4$, we find a first-order phase transition between pVBS and cVBS orders.
As shown in Fig.~\ref{fig:histogramSU6} (a), we plot the histogram of $P_\text{VBS}$ for $L=24$ at $U/t=65, 67, 67.4$ and $69$.
It is clear that the pVBS order is developed at $U/t=65$ and $67$ with the $\text{arg}(P_{\text{VBS}})$ distributed at the angles $(\frac{\pi}{4}, \frac{3 \pi}{4}, \frac{5 \pi}{4}, \frac{7 \pi}{4})$, while, cVBS order at $U/t=69$ with $\text{arg}(P_{\text{VBS}})$ distributed at $(0, \frac{\pi}{2}, \pi, \frac{3 \pi}{2})$. At $U/t=67.4$, we find the signature of the coexistence of pVBS and cVBS orders, that there are bright spots on the all the eight angles. 

To make things even clearer, we further plot the integrated distribution probability of arguments of $P_\text{VBS}$ in Fig.~\ref{fig:histogramSU6} (b). One can clearly see that there are four peaks at either the pVBS side ($U/t=65,67$) or the cVBS side ($U/t=69$), but eight peaks at $U/t=67.4$ signifying the phase coexistence~\cite{guangyuEmergent2021}, which proves that the phase transition between pVBS and cVBS is first-order.

\begin{figure}[htpb!]
	\includegraphics[width=\columnwidth]{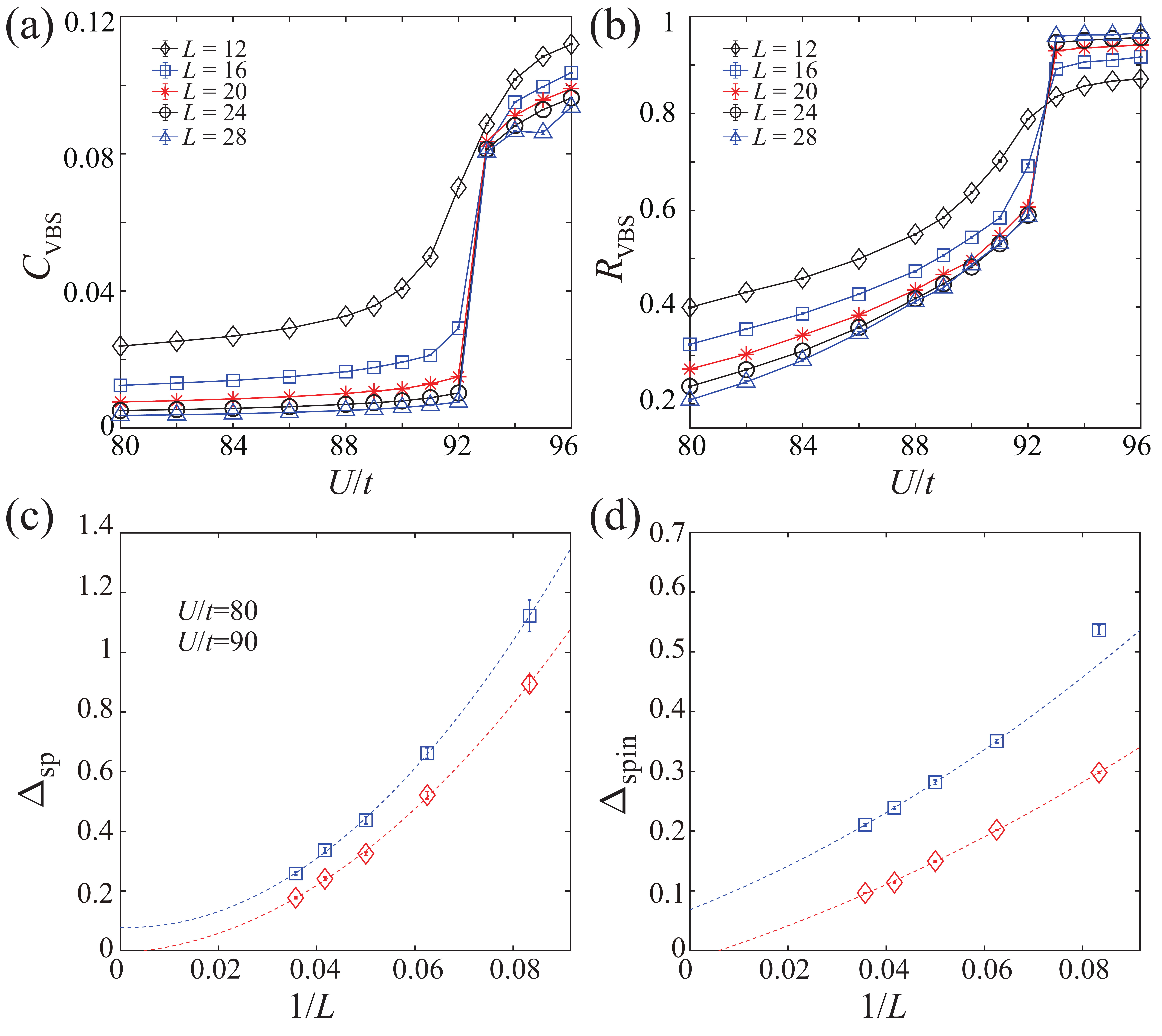}
	\caption{The (a) $C_\text{VBS}$ and (b) $R_\text{VBS}$ for different $L$ as function of $U$ in $SU(8)$ case. The DSM-pVBS GN-XY QCP is too weak to see (actually only $L=24$ and 28 data crosses each other at $U_{c1}/t\sim 90$ in (b)), but the first order transition between pVBS and cVBS is very clear at $U_{c2}/t \sim 93$. 
The $1/L$ extrapolation of (c) $\Delta_\text{sp}$ and (d) $\Delta_\text{spin}$ in $SU(8)$ case. 
We use 4 points, $L=16,20,24$ and $28$, to extrapolate $\Delta_\text{sp}$ and $\Delta_\text{spin}$ with quadratic function.
$\Delta_\text{sp}$ and $\Delta_\text{spin}$ go to $0$ at $U/t=80$, while a finite value at $U/t=90$.
Thus the DSM-pVBS transition happens between $U/t=80$ and $90$.
	}
	\label{fig:jump}
\end{figure}

\subsection{Phase diagram for $SU(8)$: DSM-pVBS-cVBS}
The similar (2+1)D GN chiral XY criticality between the DSM and pVBS and the first order transition between pVBS and cVBS as $U$ increases, also exist in the $SU(8)$ case. As shown in Fig.~\ref{fig:jump} (a) and (b), we also plot $C_\text{VBS}$  and $R_\text{VBS}$ as a function of $U$, respectively.
Since the pVBS order in this case is very weak and the parameter window is very narrow, there is no clear crossing behavior in $R_\text{VBS}$ as a function of $U$ with the system sizes accessed (actually, there is a mild cross between $L=24$ and 28, and it implies that we need even larger system sizes to clearly identify the QCP). Further increase $U$, we do find that there is a distinct jump at $U/t\sim 93$ and this is the first order transition between the pVBS and cVBS.

The single particle gap $\Delta_\text{sp}$ (Fig.~\ref{fig:jump} (c)) and the spin excitation gap $\Delta_\text{spin}$ (Fig.~\ref{fig:jump} (d)) reveal the same picture. Both are zero at $U/t=80$, but become finite at $U/t=90$, before the first-order phase transition, which means the system stays in an insulator phase at $U/t=90$.
This again means that the Mott transition (mostly likely still the GN chiral XY QCP) between the DSM and pVBS insulator happens firstly, with a relative (as compared with the $SU(6)$ case) weak and narrow pVBS phase, and the systems quickly evolve into the cVBS phase with a clear first order transition. 

\begin{figure}[htpb!]
\includegraphics[width=\columnwidth]{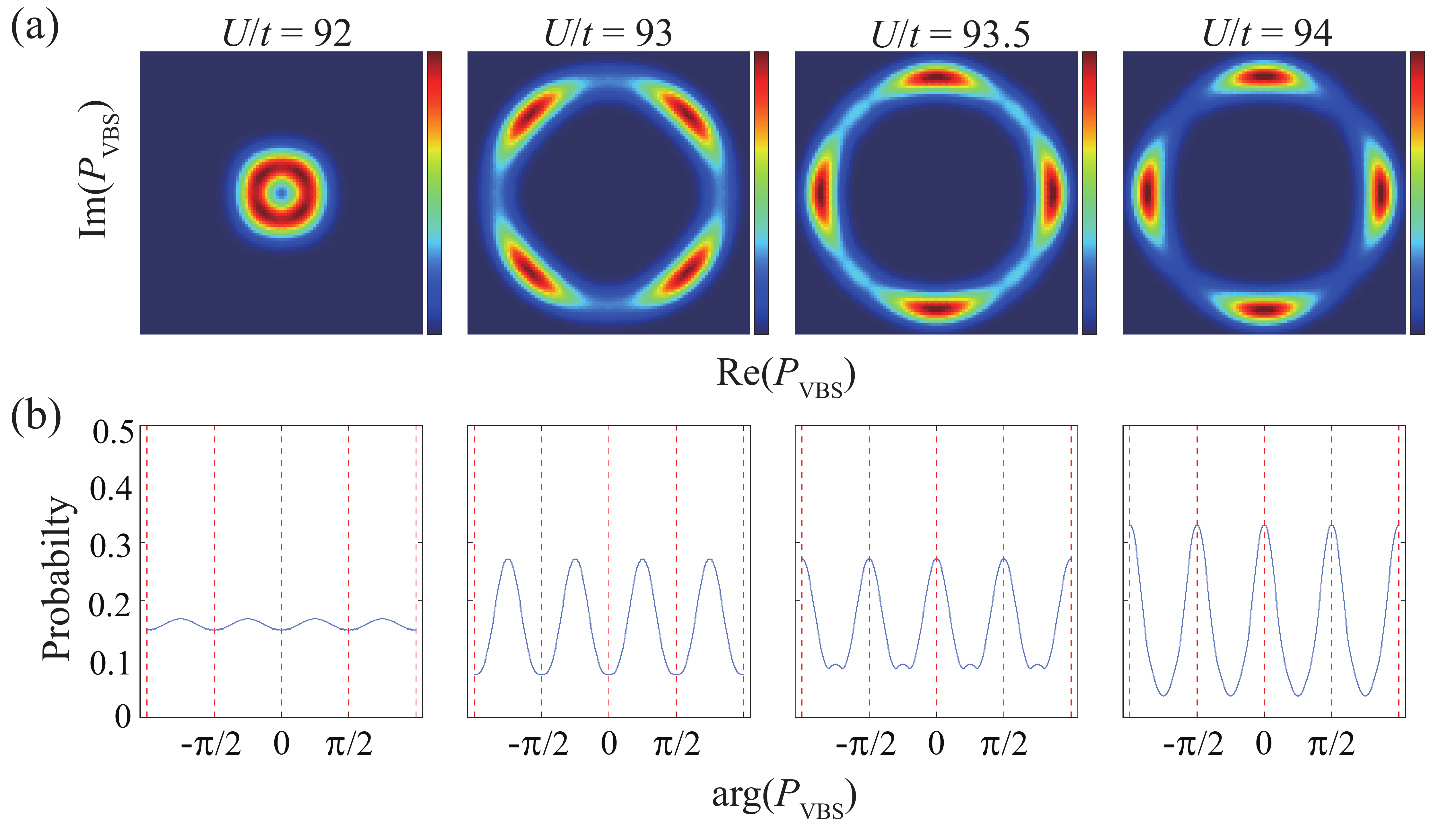}
\caption{(a) Histogram of $P_\text{VBS}$ and (b) the integrated distribution probability of arguments of $P_\text{VBS}$ in $SU(8)$ case. The color bars show the counts of $P_\text{VBS}$ data points. The transition from pVBS ($U/t=92,93$) and cVBS ($U/t=94$) is first order with phase coexistence ($U/t=93.5$) at the transition.}
	\label{fig:histogramSU8}
\end{figure}

Such an understanding is further confirmed by the histogram analysis. As shown in Fig.~\ref{fig:histogramSU8} (a) and (b), we plot the VBS order parameter histogram of $P_\text{VBS}$ for $L=24$ at $U/t=92, 93, 93.5$ and $94$. The pVBS angle distributions are seen for $U/t=92,93$ and the cVBS angle distribution is seen for $U/t=94$. And at $U/t=93.5$ both in Fig.~\ref{fig:histogramSU8} (a) and (b), the clear eight peaks stemming from the coexistence of pVBS and cVBS orders are seen. This is a robust evidence for the first order phase transition between pVBS and cVBS.

Putting together the data in Figs.~\ref{fig:jump} and ~\ref{fig:histogramSU8}, we can conclude for the $SU(8)$ case with the system sizes upto $L=28$, that, the DSM-pVBS QCP with GN chiral XY universality is $U_{c1}/t \sim 90$ and the pVBS-cVBS first order transition is $U_{c2}/t\sim 93$. Of course, the future QMC simulation data with even larger system sizes and finer parameter grid would give better estimations of the position of these quantum phase transitions.

\section{Discussions}
\label{sec:iv}
As discussed in the introduction of this paper. From the emerging research trend in 2D quantum (moir\'e) materials, the kagome metal, the ultra-cold atomic gases (including the Rydberg atom arrays), etc, it is anticipated that, by assigning more degrees of freedom to the Dirac fermions and with the extended interaction beyond the onsite Hubbard, the system will acquire larger parameter space and exhibit more interesting behavior~\cite{assaadPhase2005,Paramekanti2007sun,Cai2013quantum,Cai2013pomeranchuk,zhouMott2016,liFermion-induced2017,xuKekule2018,zhouMott2018,liaoValence2019,liuSuperconductivity2019}. Experimentally, the fermionic alkaline cold-atom arrays could realize the $SU(N)$ group with $N$ upto 10 and magnetism and Mott transitions are reported~\cite{gorshkovTwo2010,miguelUltracold2014}, and in the Rydberg atom arrays on the kagome lattice, the topological ordered phases with emergent gauge structure are reported~\cite{semeghiniProbing2021,Roushan21,Samajdar:2020hsw,yanTriangular2022}. Electrons in TBG, TMD and other quantum moir\'e material and in kagome metals are naturally bestowed with more degrees of freedom such as layer, valley and correlated flat-bands and subject to the extended and even truly long-range Coulomb interactions with high tunability~\cite{tramblyLocalization2010,tramblyNumerical2012,bistritzerMoire2011,Santos2012,lopesGraphene2007,caoUnconventional2018,shenCorrelated2020,xieSpectroscopic2019,KhalafCharged2021,KevinStrongly2020,pierceUnconventional2021,caoCorrelated2018,sharpeEmergent2019,serlinIntrinsic2020,xuKekule2018,liaoValence2019,liaoCorrelated2021,luSuperconductors2019,moriyamaObservation2019,chenTunable2020,rozhkovElectronic2016,ChatterjeeSkyrmion2020,kerelskyMaximized2019,rozenEntropic2021,tomarkenElectronic2019,soejimaEfficient2020,liuSpectroscopy2021,KhalafSoftmodes2020,ZondinerCascade2020,saitoPomeranchuk2021,GhiottoCriticality2021,SchindlerTrion2022,WangTMD2020,Parkchern2021,liaoCorrelation2021,anInteraction2020,huangGiant2020,liLattice2021,panSport2022,panThermodynamic2022,zhangSuperconductivity2021,panDynamical2022,zhangMomentum2021,liQuantum2021}. To precisely model and unbiasely solve these interesting yet difficult systems is an extremely challenging task and we believe it is in their gradually analytic and numeric solutions lie the future of the condensed matter and quantum material research. This work, in which we explore the complete phase diagram of the $SU(N)$ correlated Dirac cones subjected to the extended plaquette interaction, can be viewed as the beginning of our efforts where one generic class of interacting fermion models with extended interaction can be solved exactly.

From the large-scale quantum Monte Carlo simulations, we map out the rich phase digram of the model and find there exhibit a plethora of emerging quantum phases  such as the Dirac semimetal, the antiferromagnetic Mott insulator, interaction-driven columnar and plaquette VBS and the Dirac spin liquid and phase transitions including the Gross-Neveu chiral transitions with emergent continuous symmetry, the deconfined quantum criticality and the first order transition between different VBS states. These rich phenomena coming from the simple-looking lattice model, successfully convey the message that the interplay between $SU(N)$ Dirac fermions -- with enhanced internal symmetries and nontrival topological metric and the extended interactions beyond the Hubbard type, can indeed become the new playground to synthesise novel highly entangled quantum matter. The more realistic model design and computation solutions for the aforementioned experimental systems, are expected.

\section*{Acknowledgements}
We thank Zheng Yan for the helpful discussion on the order parameter histogram analysis. Y.D.L. acknowledges the support of Project funded by China Postdoctoral Science Foundation through Grants No. 2021M700857 and No. 2021TQ0076.
X.Y.X. is sponsored by the National Key R\&D Program of China (Grant No. 2021YFA1401400), Shanghai Pujiang Program under Grant No. 21PJ1407200, Yangyang Development Fund, and startup funds from SJTU. \
Z.Y.M. acknowledges the support from the Research Grants Council of Hong Kong SAR of China (Grant Nos. 17303019, 17301420, 17301721, AoE/P-701/20 and 17309822), the GD-NSF (no.2022A1515011007), the K. C. Wong Education Foundation (Grant No. GJTD-2020-01) and the Seed Funding “Quantum-Inspired explainable-AI” at the HKU-TCL Joint Research Centre for Artificial Intelligence.
Y.Q. acknowledges support from the the National Natural Science Foundation of China (Grant Nos. 11874115 and 12174068).
The authors also acknowledge \href{https://www.paratera.com/}{Beijng PARATERA Tech Co.,Ltd.} for providing HPC resources that have contributed to the research results reported in this paper.

\bibliographystyle{apsrev4-2}
\bibliography{main}

\end{document}